\documentclass[11pt,preprint,graphicx]{aastex}

\def\etal{\it et al. \rm }

\begin{document} 

\title{Data Mining the University: College GPA Predictions from SAT Scores}

\author{Stephen D.H. Hsu}
\affil{Department of Physics, University of Oregon, Eugene, OR 97403;
hsu@uoregon.edu}

\author{James Schombert}
\affil{Department of Physics, University of Oregon, Eugene, OR 97403;
jschombe@uoregon.edu}

\begin{abstract}

We analyze a data set comprised of academic records of undergraduates at
the University of Oregon from 2000-2004. We find correlations of roughly
0.35 to 0.5 between SAT scores and upper division, in-major GPA
(henceforth, GPA). Interestingly, low SAT scores do not preclude high
performance in most majors. That is, the distribution of SAT scores after
conditioning on high GPA (e.g., 3.5 or even 4.0) typically extends below
1000 (the average among test takers). We hypothesize that {\it
overachievers} overcome cognitive deficits through hard work, and discuss
to what extent they can be identified from high school records. Only a few
majors seem to exhibit a {\it cognitive threshold} -- such that high GPA
(mastery of the subject matter) is very unlikely below a certain SAT
threshold (i.e., no matter how dedicated or hard working the student). Our
results suggest that almost any student admitted to university can achieve
academic success, if they work hard enough.

In addition to our primary result, we find that the best predictor of GPA
is a roughly equally weighted sum of SAT and high school GPA, measured in
standard deviation units. Using a sub-population of honors college
students, we can estimate how students at elite universities would fare at
a typical state university, allowing us to comment on issues such as grade
inflation. Finally, we observe that 1) SAT scores fluctuate little on
retest (very high reliability), 2) SAT and GRE scores (where available)
correlate at roughly 0.75 (consistent with the notion that both tests
measure a stable general cognitive ability) and 3) the SAT distribution of
students that obtained a degree does not differ substantially from that of
the entering class.

\end{abstract}

\section{Introduction and Overview} 

Considering their widespread use in college admissions, the predictive
power (validity) of tests such as the SAT and ACT is a surprisingly
controversial topic. On the negative side, one often reads claims that the
correlation of SAT with freshman GPA is as low as 0.25 to 0.35
(Sackett \etal 2008). On the positive side, one cannot help but be impressed
by a measure of cognitive ability that requires only a few hours of
testing, is fairly stable (see results below) and has roughly as much
predictive power as high school GPA, which represents years of evaluation
by experienced instructors\footnote{For a review of psychometrics and the
science of mental ability, see Jensen (1998). The SAT is clearly a
cognitive test, and as such has a high correlation with IQ or the general
factor of intelligence, $g$. Indeed, the SAT, administered at an early age
to overcome ceiling limitations, has been used successfully as a tool for
identifying exceptionally gifted children (Park \etal 2008).}. Note, in what
follows, we will refer primarily to SAT rather than SAT/ACT, and convert
scores when necessary to the SAT scoring system.

It is easy to understand why freshman GPA measured over an entire
population is not a satisfactory metric of academic success -- students
typically self-select into courses of varying difficulty already in their
freshman year. More able students typically take more difficult courses,
for example: linear algebra, multivariable calculus or electromagnetism,
whereas less able students are likely to be in introductory courses that
are not very different from high school classes. A further complication is
that studies done at elite universities suffer from restricted range:
nearly all students in such studies have high SAT scores. Theoretical
estimates suggest that adjusting for effects such as course difficulty and
restriction of range leads to higher correlation values such as 0.55
(Berry \& Sackett 2009).

The University of Oregon (UO), a state flagship campus, has a number of
favorable characteristics for the study of SAT predictive power. Because UO
admits a broad range of students, there is little restriction of range:
very strong, average and slightly below average (i.e., below the average
among all test takers, SAT $<$ 1000) students are represented. The student
body also has a fairly uniform racial composition, which minimizes effects
of ethnicity on the results.

In the analysis presented below, we focus on upper division, in-major GPA.
This corrects for variation between majors in course difficulty and grading
standards. Upper division grades test in-depth mastery of material specific
to each major. While we do observe variation between different majors,
which we explore in some detail, in almost every major we find a
significant correlation between SAT and GPA, with values ranging from about
0.35 to 0.5.

Thus, at least for UO and similar institutions, a strong case can be made
for the use of SAT scores in admissions. Students with high SAT scores are
clearly more likely to perform well in upper division courses. However, it
is important to emphasize that a low SAT score does not, by itself,
preclude strong performance, a fact that surprised us when we examined the
data. There are many students (overachievers) with modest scores who
nevertheless achieve high upper division GPAs, across a broad variety of
majors. Interestingly, these overachievers tend to be disproportionately
female (64 percent), whereas underachievers (students with high SAT scores
but poor performance in courses) are overwhelmingly male (79 percent).

Our results raise a number of interesting questions, which we investigate
further below. Can one determine from high school records which low scoring
students are likely to overachieve?  Perhaps work ethic or
conscientiousness in low SAT students is reflected in superior high school
grades? Some studies (Noftle \& Robins 2007) have shown that conscientiousness (as
measured in personality inventory) contributes to college GPA even after
high school GPA$_{HS}$ and SAT are controlled for. Below, we determine the
relative weighting of GPA$_{HS}$ and SAT that best predicts upper division
GPA, across a variety of majors.  A related analysis, using data from the
Clark Honors College at UO, allows us to predict the performance of elite
university students were they to have enrolled at a typical state
university instead.

A final question relates to whether mastery of certain subjects is more
cognitively challenging than others. Such majors might exhibit {\it
cognitive thresholds} below which no amount of hard work or motivation is
enough to permit mastery of the subject matter. In section \S4 we discuss
some modest evidence for cognitive thresholds related to mathematical
ability in majors such as physics and math.

\section{Descriptive Results} 

Our dataset focuses on courses taken from 2000 to 2004 by students in
twelve majors in the College of Arts and Sciences (see below).  In order to
follow the academic path from freshman year to graduation, we constrained
the sample to only include those students enrolled during the 2000 to 2004
time frame, admitted after 1996 (to avoid the 1995 SAT re-centering event).
Students were then binned into four types; 1) those who graduated in one of
the twelve majors selected for study, 2) those who graduated, but changed
their major to something outside the selected twelve, 3) those who
graduated, but outside a six year limit, and 4) those who did not graduate
(this will include those who dropped out or transferred or are simply still
working on their degree).  Note that by limiting the sample to students
enrolled before or during 2004, we include all students who should have
graduated in less than six years by 2010.

\begin{figure}
\centering
\includegraphics[scale=0.6]{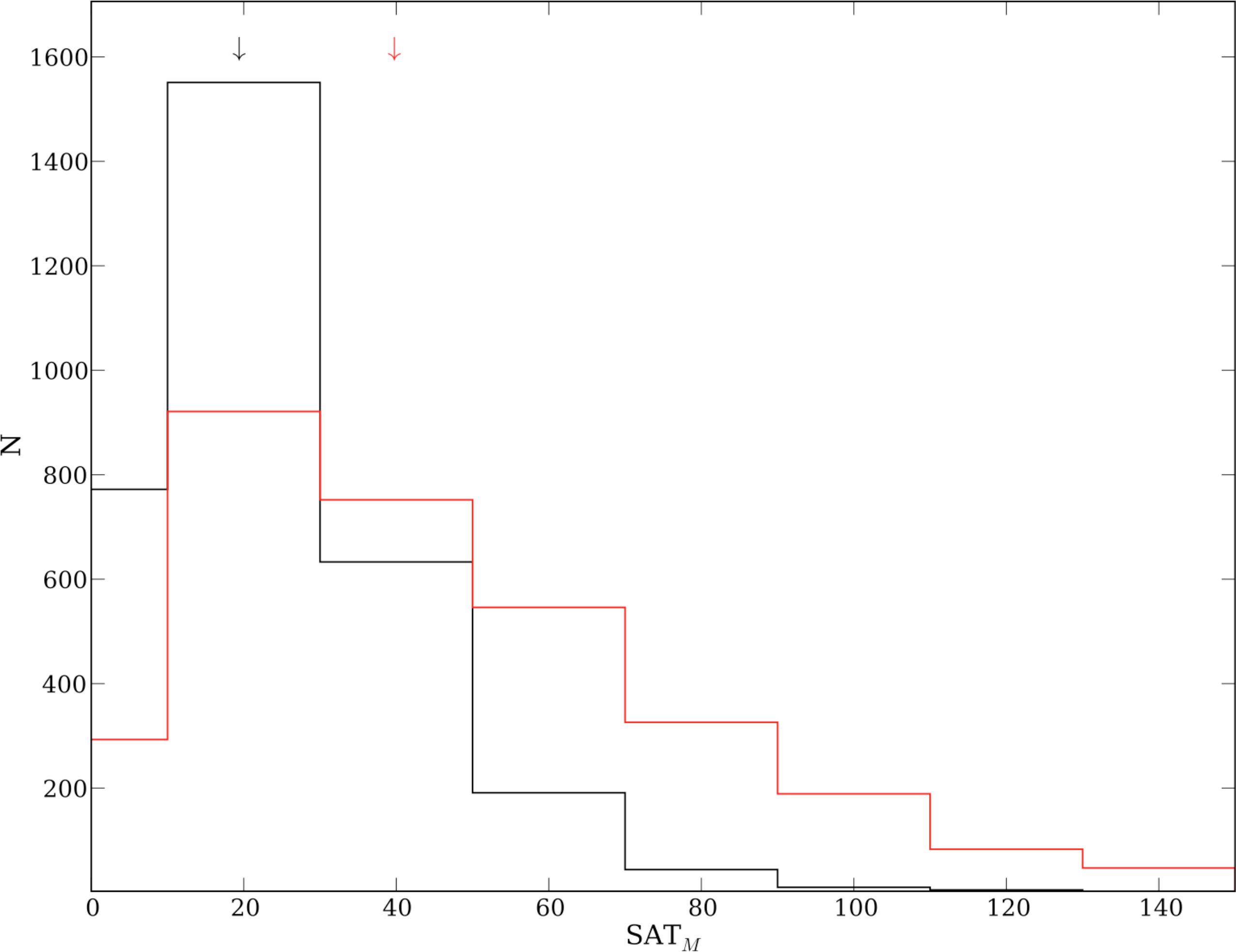}
\caption{Histogram of SAT repeat scores for the math section.  The black
histogram represents the frequency of occurrence of the difference
between the maximum score and the average score for given individuals.  The
red histogram displays the difference between the minimum score and maximum
score.  The mean difference between average and maximum is only 20 points.}
\end{figure}

For uniformity in the sample, we selected twelve majors from the College of
Arts and Sciences (CAS).  They are: Biology, Chemistry, Computer and
Information Sciences, Economics, English, History, Mathematics, Philosophy,
Physics, Political Science, Sociology and Spanish.  The total sample
contained 9,934 students of which 2,474 did not graduate and 746 took more
than six years to obtain their degree.  The resulting analysis sample
contains 6,714 students (although 2,249 lacked SAT or ACT scores, or
graduated in some major outside the twelve selected for analysis).  The
final sample contained 4,420 students over all twelve majors.  Of those,
only 336 lacked high school GPAs, producing a total sample size of 4,084
students.  Table 1 lists the total numbers for the sample by major, mean
GPAs and SAT scores.

\begin{deluxetable}{rrcccccccc}
\tablecolumns{11}
\small
\tablewidth{0pt}
\tablecaption{Mean GPAs and SAT scores for twelve CAS majors}
\tablehead{
\colhead{Major} &
\colhead{N} &
\colhead{GPA$_{total}$} &
\colhead{GPA$_{upper}$} &
\colhead{SAT$_{M}$} &
\colhead{SAT$_{R}$} &
\colhead{SAT$_{M+R}$} &
\colhead{GPA$_{HS}$} \\
}
\startdata

All Majors        & 4420 & 3.20$\pm$0.50 & 3.27$\pm$0.53 & 556$\pm$85 & 572$\pm$90 & 1128$\pm$155 & 3.45$\pm$0.43 \\
Biology           &  521 & 3.28$\pm$0.44 & 3.36$\pm$0.47 & 597$\pm$74 & 590$\pm$82 & 1188$\pm$133 & 3.65$\pm$0.38 \\
Chemistry         &   84 & 3.35$\pm$0.42 & 3.21$\pm$0.50 & 623$\pm$77 & 610$\pm$88 & 1232$\pm$147 & 3.79$\pm$0.26 \\
CIS               &   25 & 3.29$\pm$0.50 & 3.41$\pm$0.59 & 635$\pm$59 & 590$\pm$90 & 1225$\pm$119 & 3.61$\pm$0.30 \\
Economics         &  445 & 2.95$\pm$0.54 & 2.97$\pm$0.63 & 569$\pm$81 & 530$\pm$98 & 1099$\pm$157 & 3.35$\pm$0.44 \\
English           &  572 & 3.28$\pm$0.46 & 3.32$\pm$0.46 & 552$\pm$79 & 608$\pm$81 & 1160$\pm$141 & 3.48$\pm$0.42 \\
History           &  500 & 3.27$\pm$0.45 & 3.29$\pm$0.45 & 548$\pm$80 & 585$\pm$79 & 1133$\pm$141 & 3.43$\pm$0.43 \\
Mathematics       &   60 & 3.48$\pm$0.49 & 3.23$\pm$0.68 & 662$\pm$62 & 641$\pm$82 & 1304$\pm$125 & 3.60$\pm$0.52 \\
Philosophy        &  164 & 3.29$\pm$0.45 & 3.36$\pm$0.44 & 577$\pm$77 & 604$\pm$84 & 1181$\pm$141 & 3.39$\pm$0.42 \\
Physics           &   77 & 3.22$\pm$0.55 & 3.16$\pm$0.56 & 647$\pm$70 & 621$\pm$77 & 1269$\pm$131 & 3.57$\pm$0.50 \\
Political Science &  728 & 3.18$\pm$0.50 & 3.21$\pm$0.55 & 543$\pm$80 & 574$\pm$83 & 1117$\pm$145 & 3.36$\pm$0.43 \\
Sociology         &  744 & 2.98$\pm$0.52 & 3.18$\pm$0.54 & 508$\pm$80 & 522$\pm$86 & 1031$\pm$147 & 3.31$\pm$0.41 \\
Spanish           &  500 & 3.45$\pm$0.40 & 3.61$\pm$0.40 & 552$\pm$79 & 572$\pm$85 & 1124$\pm$146 & 3.57$\pm$0.38 \\

\enddata
\end{deluxetable}

Student GPA was calculated in two ways. The traditional quantity,
GPA$_{total}$, uses all courses taken regardless of course title or credit
hours.  Courses marked 'P' (pass), 'N' (no pass), 'W' (withdraw), 'Y' (no
grade reported) or 'I' (incomplete) were ignored in our analysis.
GPA$_{upper}$, a measure of upper division performance in a specific
subject, was calculated using only courses in the student's major, at or
above the 300 level. Upper division GPA presumably reflects mastery of a
chosen field of study.  In the few cases where a student was a double
major, upper division GPA was calculated twice for each major and recorded
as separate data points.  The set of upper division courses for Mathematics
and CIS (Computer and Information Sciences) were restricted to a subset of
courses that are considered rigorous and are typically prerequisites for
post-graduate study.

\begin{figure}
\centering
\includegraphics[scale=0.6]{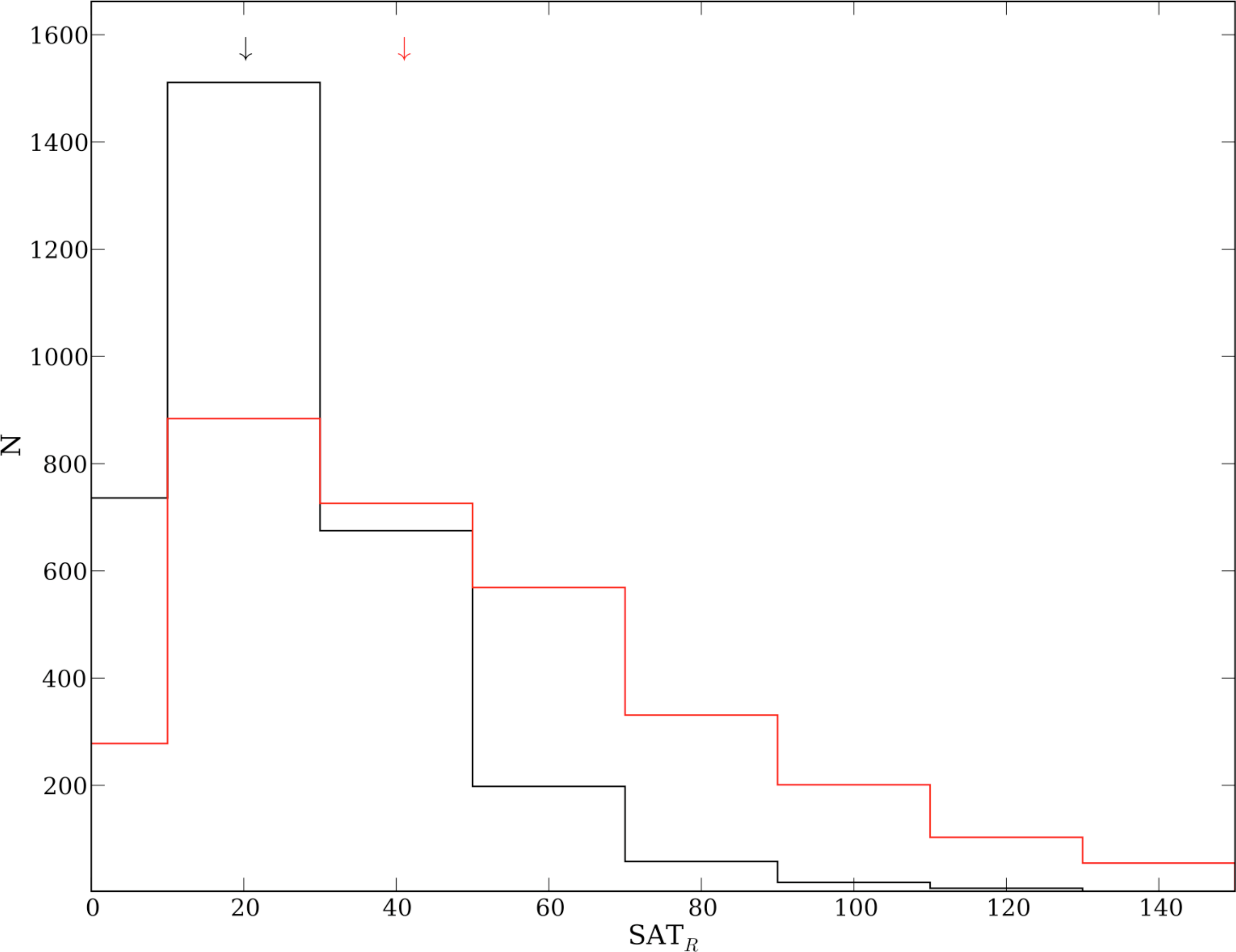}
\caption{Histogram of SAT repeat scores for the reading section.  The black
histogram represents the frequency of occurrence of the difference
between the maximum score and the average score for given individuals.  The
red histogram displays the difference between the minimum score and maximum
score.  The mean difference between average and maximum is only 20 points.}
\end{figure}

As outlined in greater detail in the appendix, the ability to perform the
analysis in this paper, particularly the separation of GPA for upper
division courses, required a combination of knowledge of student records,
access to lists of majors over various years, network tools to extract the
necessary data, software tools to interpret the data, and an understanding
of data reduction of large datasets.  Most of the analysis involved in this
study could not have been produced by simple SQL commands to various
databases supported by student record software.  In particular, typical
access to student records in the current UO network environment is through
webpages.  Analysis of large numbers of student records requires the
parsing and interpretation of HTML, generating structured data from
unstructured sources.

\subsection{SAT/ACT conversion and reliability of SAT scores}

From the final sample, 4,179 ($95\%$) had SAT scores.  The rest had ACT
scores, which we converted to SAT values using standard ACT to SAT
concordance tables. In aggregate, our sample contains 8,085 students with
SAT scores or converted ACT scores. Of these students, 2,486 took the Math
and Reading components twice, 562 took them three times, 114 took them four
times, 31 took them over 5 times.  This allows us to analyze the effect of
multiple re-tests of the SAT by looking at the mean score versus the peak
score and the minimum score versus the peak score.  This analysis is shown
in Figures 1 and 2, and demonstrates that, on average, re-taking the SAT
only improves scores by 20 to 30 points per section.  The use of average
vs. peak SAT score has little impact on our subsequent analysis.

\begin{figure}
\centering
\includegraphics[scale=0.6]{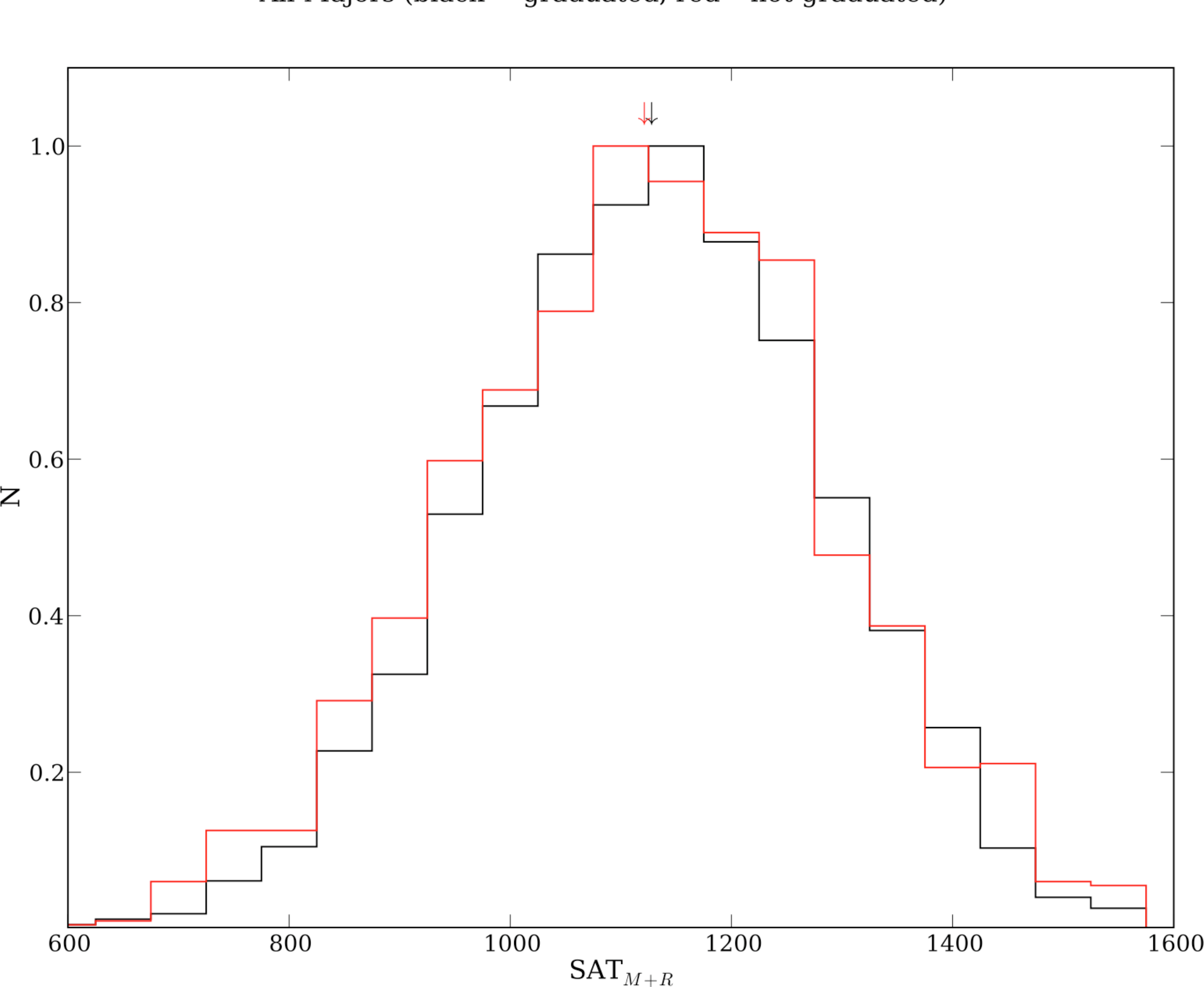}
\caption{The distribution of composite SAT scores (math plus reading) for
students in one of our fourteen majors that graduated in less than six
years, versus composite SAT scores for students that did not graduate.  The
histograms are normalized to their peak values.}
\end{figure}

In addition, from our initial search samples, we have 740 students with SAT
scores and recent GRE scores (typically graduating seniors applying for
admission to graduate programs).  The composite SAT scores range from 800
to 1550 The SAT and GRE scores are highly correlated (with R values of
roughly 0.75 for both Math and Reading) and a correlation slope near unity.
This correlation is somewhat lower than the .86 obtained in Angoff (1990),
however that study was conducted before the 1995 re-centering of the SAT.

\subsection{SAT Scores for Graduates versus Non-Graduates}

Figure 3 displays the composite SAT scores (math plus critical reading) for
students graduating from UO within one of the twelve majors (4,420
students) versus those students who did not graduate (2,474 students).
Note that this includes the entire population of students who may have
transferred to other schools with better (or different) programs, as well
as students who left college entirely. Based on overall UO attrition rates,
we suspect that most of the non-graduates actually left the university,
although we do not have precise data to support this. Interestingly, the
distribution of SAT scores for both populations is identical.  Not only do
the means agree within the errors, but also the variance and shape of the
distributions are identical.  We conclude that, as a population, students
leave college for reasons unrelated to their cognitive ability.

\begin{figure}
\centering
\includegraphics[scale=0.6]{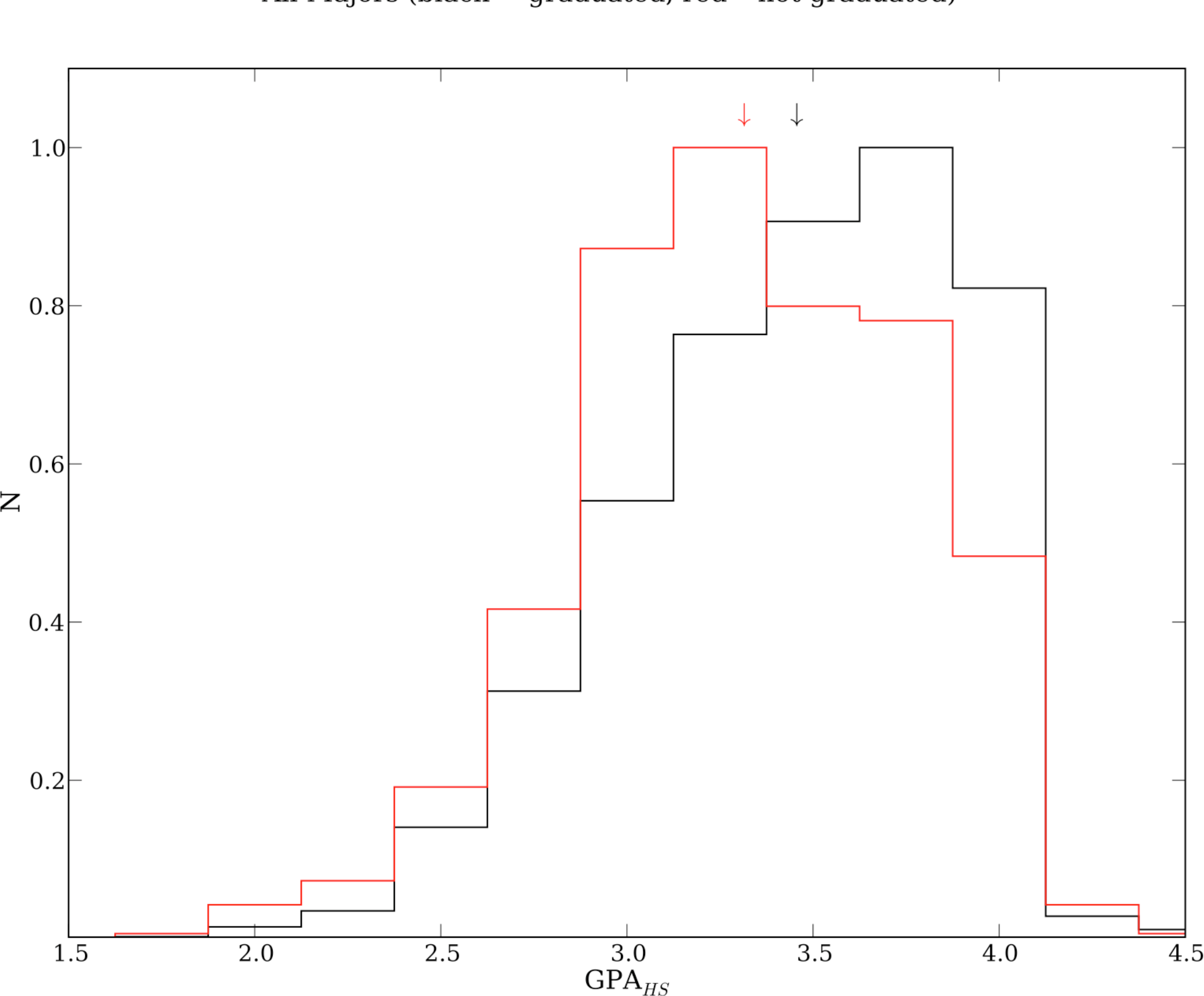}
\caption{The distribution of high school GPA (GPA$_{HS}$) for students in
one of our fourteen majors that graduated in less than six years, versus
high school GPA for students that did not graduate.  The histograms are
normalized to their peak values.}
\end{figure}

In comparison, high school GPA (GPA$_{HS}$) is a better predictor of
graduation probability.  Figure 4 displays the distribution of HS GPA for
the same populations in Figure 3.  There is a clear shift in the peaks
of the histograms, to lower GPA$_{HS}$ for the students that did not
graduate.  This may reflect a lack of ability to perform in a classroom
setting.  Retention efforts at various universities should concentrate on
students with low high school GPAs rather than SAT scores.

\section{SAT versus upper division GPA}

The core of this project is to examine the relationship between entering
student SAT score and academic success.  We measure academic success in
terms of grades in courses within a student's major.  To this end, we
define an upper GPA (GPA$_{upper}$) for in-major classes with course numbers
greater than 300 (i.e., upper division classes).  This procedure ignores
grades outside their major and any courses below the level of 300.  Our
objective is to isolate a metric which correlates with mastery of an
academic subject.  For example, most programs require a GPA of 3.8 for
selection of graduation with honors.

\begin{figure}
\centering
\includegraphics[scale=0.8]{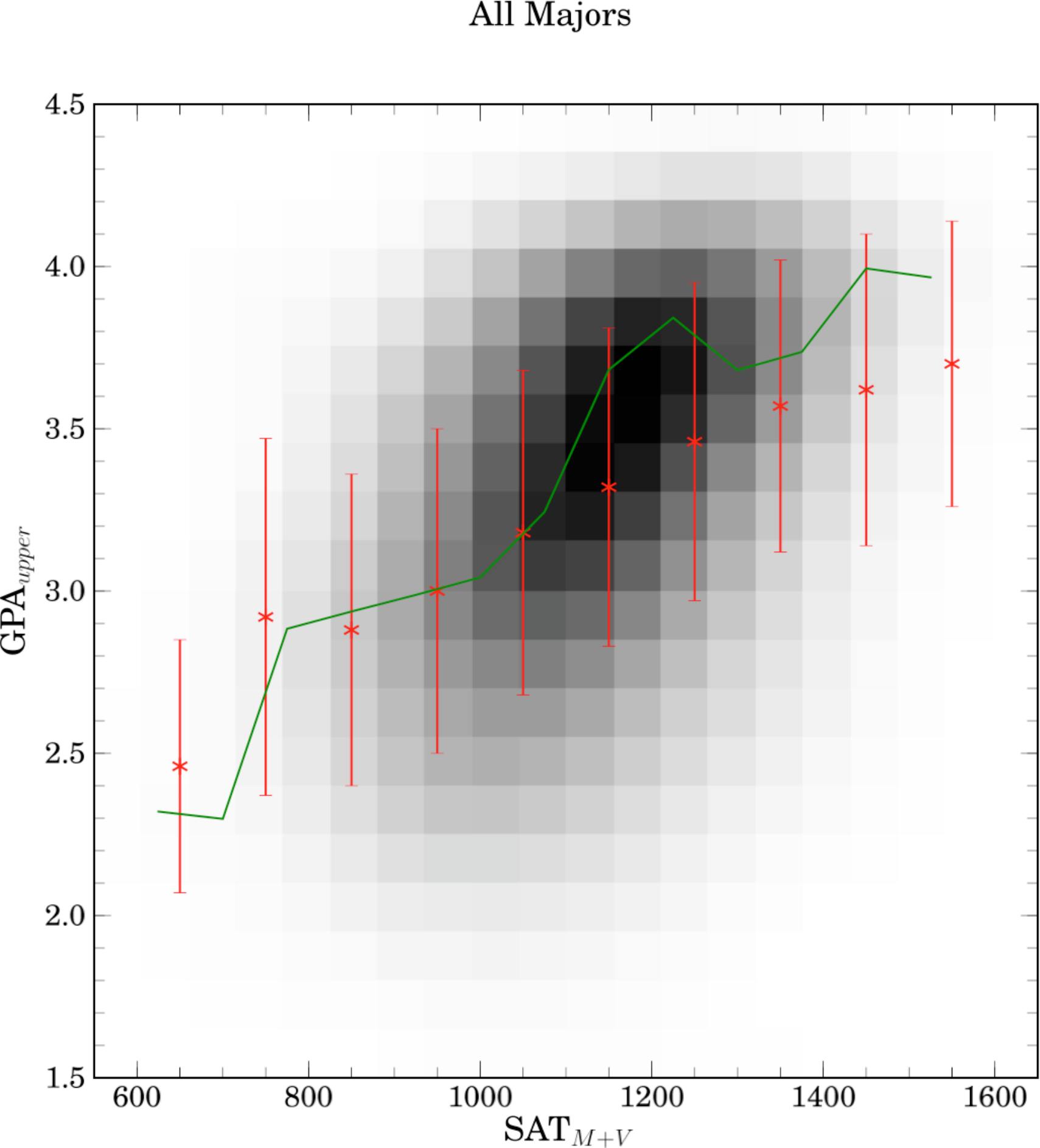}
\caption{The greyscale density distribution of all 4,420 students' combined
SAT scores versus upper division GPA.  A clear trend for higher GPA with
higher SAT scores is noted.  Also shown is the moving average (red symbols)
binned in SAT intervals of 100.  Error bars display a standard deviation on
each average (not the error).  The green line follows the peak of the
density distribution.}
\end{figure}

While we expect that upper GPA is correlated with total GPA, we cannot
ignore the fact that grade inflation, and its counterpart, diluted course
content, are particularly systemic in the lower division classes.  This
makes lower division courses a much poorer measure of academic success due
to the large range in course type, grading, content and instructor quality.

The resulting plot of SAT score versus GPA$_{upper}$ for all majors (4,420
students) is shown in Figure 5.  Due to the large number of data points,
a scatter plot is not the best choice for conveying visual information.
Instead, we plot the data as a normalized density distribution.  Each data
point is treated as a small 2D gaussian with standard deviation equal to
the size of a (small) selected grid cell.  Each grid cell has a value
determined by summing the value of all the gaussians at the cell's center
point.  This results in a greyscale plot where darker cells represent more
data.  The shading intensity of each cell provides additional visual
information beyond the position of the cell within the two variable space.

\begin{figure}
\centering
\includegraphics[scale=0.9]{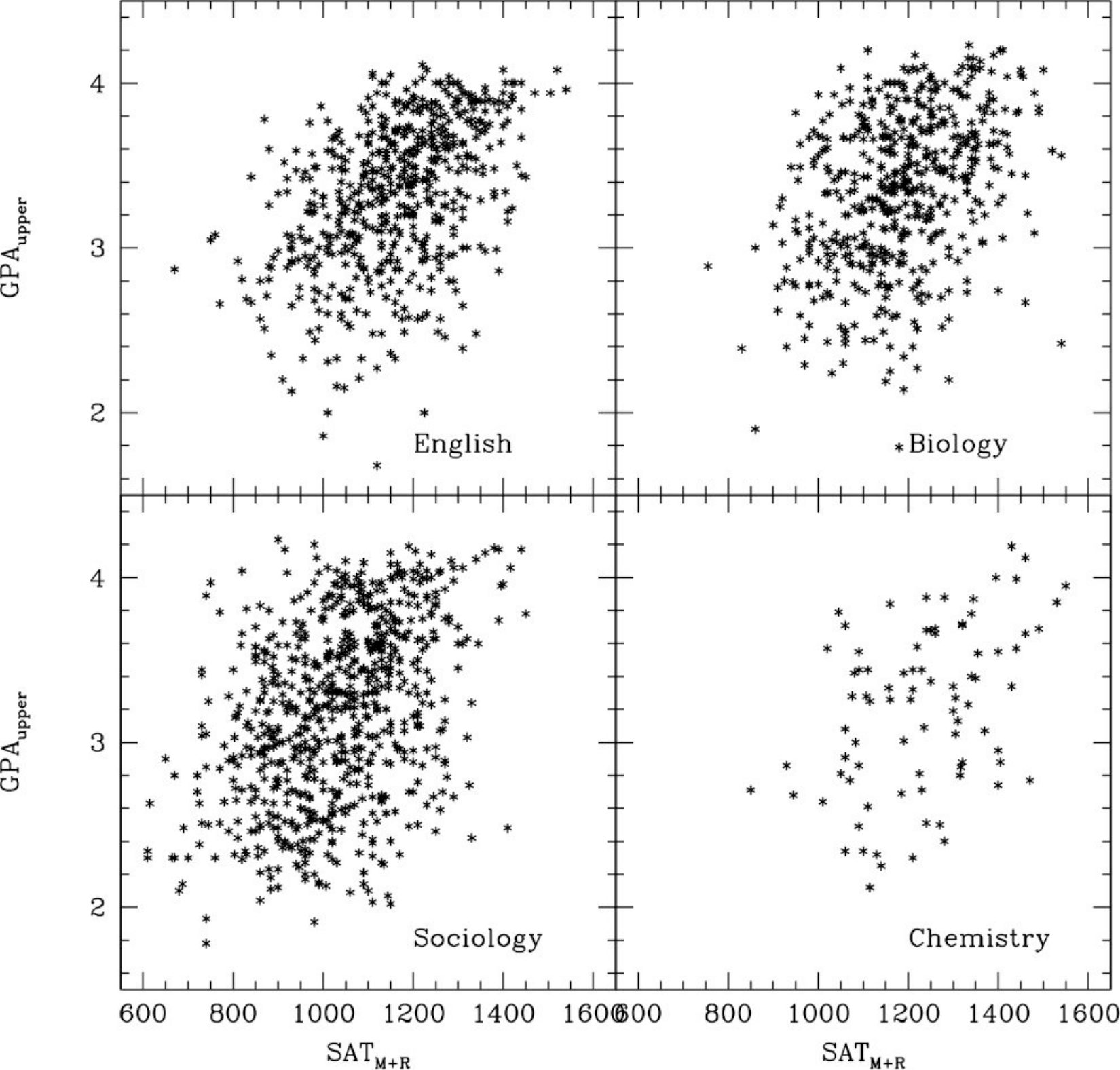}
\caption{Combined SAT scores versus upper division GPA for four majors;
English, Biology, Sociology and Chemistry.  These four are presented to
illustrate the possible distributions of scores and GPA in different
majors.  Plots of all twelve majors analyzed can be found at our data
website, http://abyss.uoregon.edu/$\sim$js/sat}
\end{figure}

The greyscale image shows the distribution of SAT versus GPA$_{upper}$
scores for our entire sample.  There is a clear trend for increasing SAT
score with increasing GPA (R=0.40).  Also displayed on this plot are the
moving averages in GPA (the red symbols and the errors bars of one standard
deviation) and the peak density positions (the green line).  All three
demonstrate an increase in GPA with larger SAT score.

The same analysis can be performed on any of the twelve majors.  All of the
relevant plots can be found at our website
(http://abyss.uoregon.edu/$\sim$js/sat). For reference we display the SAT
versus GPA plots of English, Biology, Sociology and Chemistry in Figure 6.
While the same general trends of increasing SAT score with increasing GPA
are seen in all the majors, each major exhibits subtle differences which
may reflect the nature of their various degree programs.  For example,
Sociology has more students in the upper left portion of the diagram (low
SAT score, high GPA) than other majors.  English has fewer low GPA students
than other majors.

\begin{figure}
\centering
\includegraphics[scale=0.8]{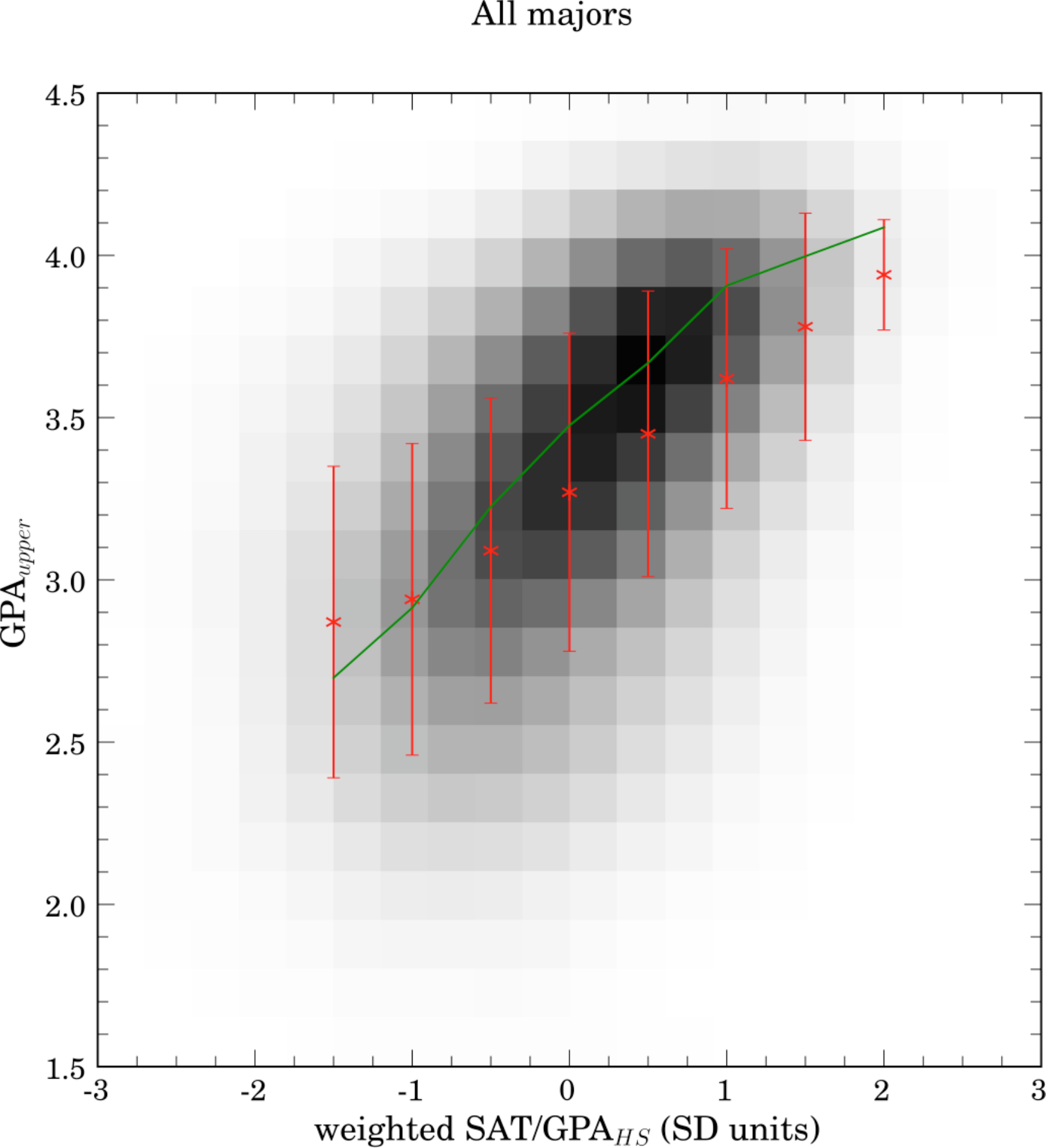}
\caption{The two component model of combined SAT scores and high school GPA
versus upper division GPA.  Again, the moving average and peak values are
shown following Figure 6.}
\end{figure}

For the sake of later discussion, we divide the SAT vs GPA diagram into
four areas; upper left and right, lower left and right.  Clearly, the upper
right and lower left portions of the diagram are populated by students
within normal expectations, i.e., they are performing as one might expect
given their SAT scores.  The upper left and lower right display populations
that, to some degree, deviate from expectations.  The lower right
represents students who struggle in college, despite having superior SAT
scores.

The upper left portion of the diagram represents the most interesting
population of students, those who over-perform in their majors relative to
the expectations from their SAT scores.  We refer to these students as
overachievers, suspecting that their high performance may be due to factors
such as conscientiousness or personal motivation.  We return to this
population later in our study.

\subsection{Best Fit Predictor: SAT + GPA$_{HS}$}

There are significant correlations between SAT score and GPA$_{upper}$ as
well as between GPA$_{HS}$ and GPA$_{upper}$. Since SAT and GPA$_{HS}$ are
not perfectly correlated, we expect that some combination of SAT score and
GPA$_{HS}$ will exhibit an even stronger correlation with GPA$_{upper}$. To
construct this optimal predictor, we first normalize SAT scores and high
school GPA to their various means and standard deviations ($z$ score).  We
then test a range of linear combinations searching for the values which
maximize the correlation against upper GPA.  These fractional values
($f_{max}$, $R_{max}$) are shown in Table 2 for each major, where the
combination is expressed as $(1 - f) \cdot {\rm SAT} + f \cdot {\rm
GPA_{HS}}$.  Table 2 also displays the correlation coefficient for
GPA$_{upper}$ versus SAT ($f = 0$, $R_{SAT}$) and the correlation
coefficient for GPA$_{upper}$ versus GPA$_{HS}$ ($f = 1$, R$_{GPA}$).

A plausible hypothesis for overachievers is that they overcome cognitive
deficiencies through conscientiousness and hard work.  We would expect
these behavioral qualities to be present already at the high school level,
and reflected in high school grades.  Thus, we expect that these same
academic behavior advantages would also carry over into college
performance.

\begin{figure}
\centering
\includegraphics[scale=0.8]{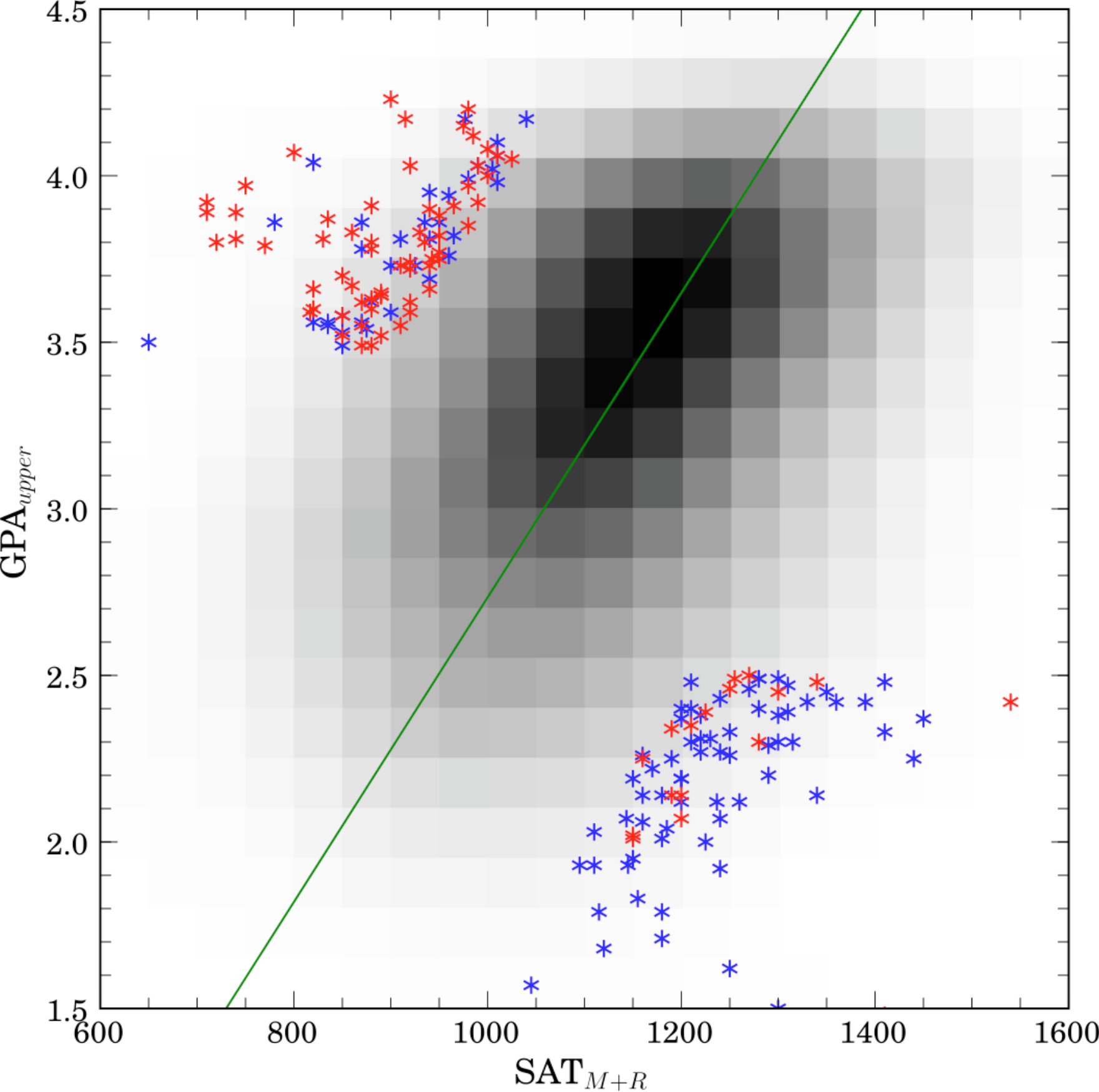}
\caption{Underachievers and Overachievers (red = females, blue = males)
isolated by SAT and upper GPA (1.25 standard deviations from the green
ridgeline).  The overachievers are mostly female students (64\%) and
the underachievers are mostly male students (79\%).}
\end{figure}

The data in Table 2 show a modest improvement in prediction of academic
success using both SAT score and high school GPA.  While there is some
variation between different majors (e.g., Philosophy versus Sociology), the
best predictor is typically obtained for nearly equal weighting of SAT and
high school GPA.  This two component model is shown graphically in Figure
7.

\begin{deluxetable}{rrcccccccc}
\tablecolumns{11}
\small
\tablewidth{0pt}
\tablecaption{Correlation Coefficients for fits to GPA and SAT scores}
\tablehead{
\colhead{Major} &
\colhead{$R_{SAT}$} &
\colhead{$R_{GPA}$} &
\colhead{$f_{max}$} &
\colhead{$R_{max}$} \\
}
\startdata

All Majors        & 0.40 & 0.46 & 0.61 & 0.52 \\
Biology           & 0.36 & 0.40 & 0.61 & 0.47 \\
Chemistry         & 0.36 & 0.40 & 0.52 & 0.45 \\
CIS               & 0.00 & -0.09 & 0.11 & -0.01 \\
Economics         & 0.49 & 0.43 & 0.44 & 0.57 \\
English           & 0.44 & 0.46 & 0.52 & 0.57 \\
History           & 0.36 & 0.43 & 0.61 & 0.48 \\
Mathematics       & 0.46 & 0.27 & 0.31 & 0.48 \\
Philosophy        & 0.42 & 0.29 & 0.31 & 0.44 \\
Physics           & 0.40 & 0.37 & 0.51 & 0.48 \\
Political Science & 0.43 & 0.46 & 0.53 & 0.54 \\
Sociology         & 0.40 & 0.54 & 0.71 & 0.58 \\
Spanish           & 0.45 & 0.48 & 0.54 & 0.56 \\

\enddata
\end{deluxetable}

\subsection{Underachievers and Overachievers}

We previously defined overachievers as students who entered the university
with low SAT scores, but earned high upper division GPAs. Similarly, the
underachievers had high SAT scores but performed poorly in college. We
further refine our definition of these groups as follows. In Figure 8,
which displays upper GPA versus combined SAT, we fit a line to the ridge of
highest density (green curve). We then calculate the standard deviation of
population density distribution in the direction perpendicular to this
line. Students that lie more than +1.25 SD above the line (with GPAs
greater than 3.5) are defined as overachievers, and those who are more than
-1.25 SD below the line (with GPAs less than 2.5) are defined as
underachievers.  The outcome of this procedure is displayed in Figure 8, in
which the overachievers are in the upper left (red are female students,
blue are male students) and the underachievers are in the lower right.

An analysis of the two populations reveals that overachievers are
predominantly (64 percent) female and underachievers are overwhelmingly (79
percent) male. See Figure 8 for gender distributions; note the overall
population of our data set has a male:female ratio of 45:55.  Table 3 and 4
display the breakdown by major and gender for the over and underachievers.
Social sciences dominate both categories, although one should be careful in
considering the distribution by major. We defined over- and underachievers relative
to the GPA vs. SAT trend computed across all 12 majors. Some majors, whose average
upper GPAs deviate from the group average, will tend to overpopulate one of the two groups. 
For example, Spanish has a high average upper GPA (3.61) and economics has a low one (2.97). 
This contributes to the overrepresentation of Spanish majors among the overachievers, and of economics majors among the underachievers. If our primary interest were the distribution
among majors, we could correct for these variations in average upper GPA by converting to SD units relative to mean for each major. Note that Sociology, which contributes the single largest group of overachievers, has an average upper GPA of 3.18, which is close to the average of 3.20 for the 12 majors as a whole. Thus, in the case of Sociology, the overrepresentation of overachievers is not due to systematically higher grades.

One might guess that these two groups could be identified using high school
GPA. Figure 9, in which the horizontal axis is GPA$_{HS}$, shows a wide
range for both populations. Many of the underachievers have low high school
GPA relative to what their SAT scores would have predicted, and lower than
the average for admitted UO students. Figure 10 shows that women tend to outperform men in upper GPA at any fixed value of SAT score. This is related to the fact that women admitted to the University tend to have higher high school GPAs. In the best fit predictive model we discussed above, women would have higher combined (SAT + HSGPA) z scores than men at fixed values of SAT score.

\begin{figure}
\centering 
\includegraphics[scale=0.8]{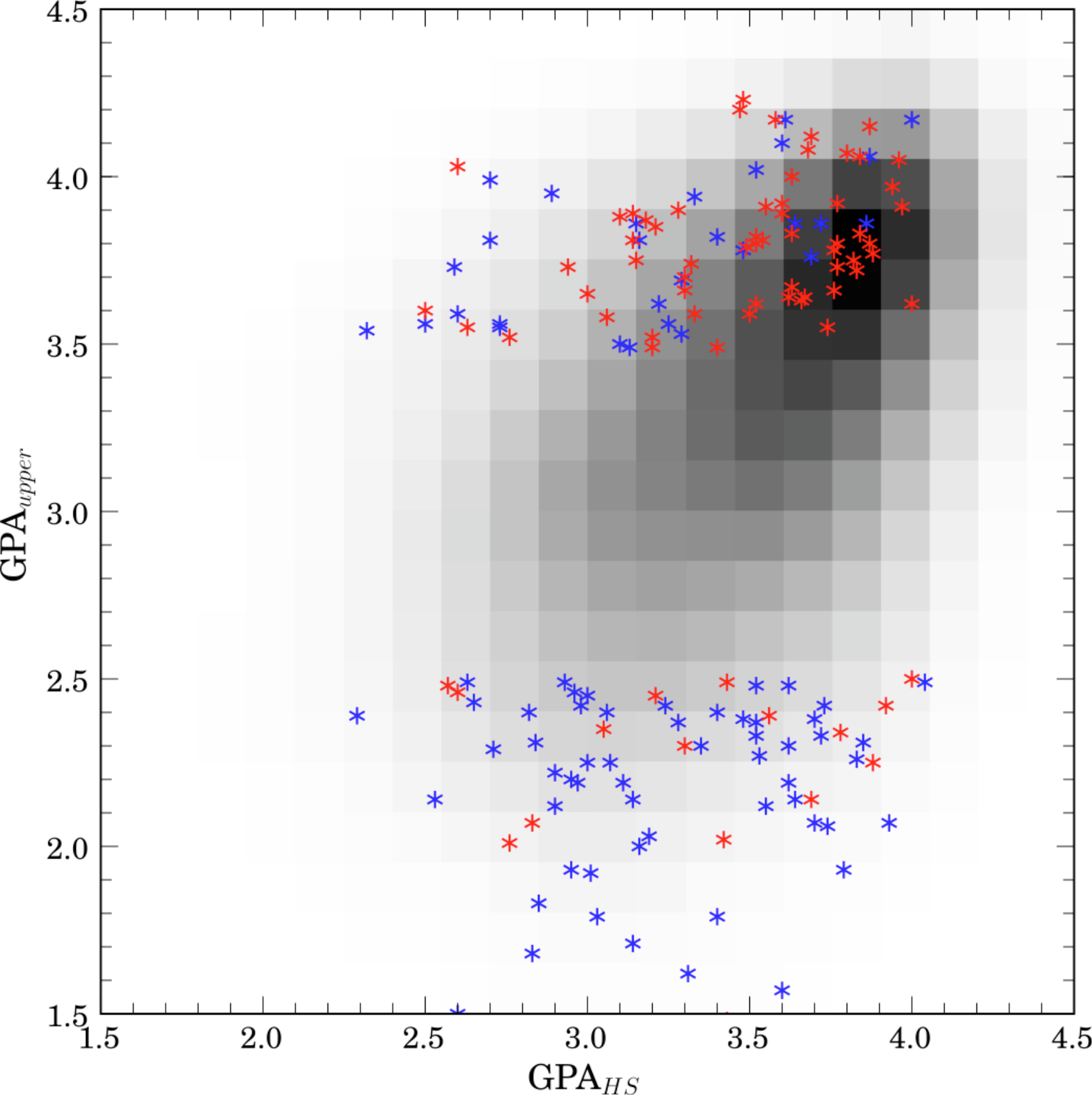}
\caption{Underachievers and Overachievers (red = females, blue = males)
from the previous diagram, now plotted by their high school GPA. Both
overachievers and underachievers
display a wide range of high school GPAs.}
\end{figure}

\begin{figure}
\centering
\includegraphics[scale=0.6]{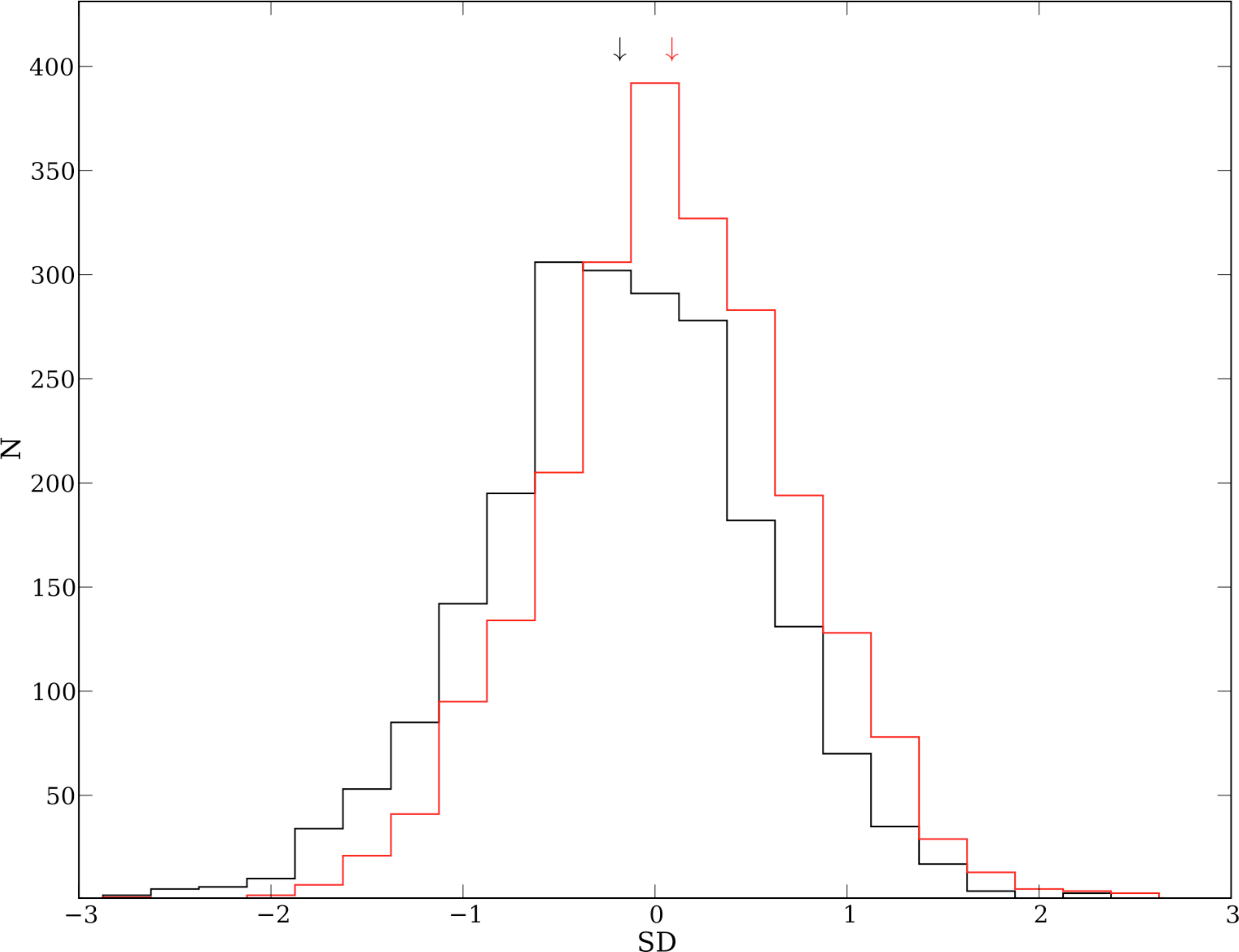}
\caption{Histogram of standard deviation (SD) from the GPA/SAT ridgeline
(red = females, black = males).  Notice that female students outperform
male students over the full range of SAT score or upper GPA.}
\end{figure}

\begin{deluxetable}{rlrlrl}
\tablecolumns{6}
\small
\tablewidth{0pt}
\tablecaption{Overachievers}
\tablehead{
\colhead{} &
\colhead{Male} &
\colhead{} &
\colhead{Female} &
\colhead{} &
\colhead{Total} \\
}
\startdata

     &                   &  1 & Biology           &  1 & Biology \\
   3 & Economics         &  2 & Economics         &  5 & Economics \\
   1 & English           &  1 & English           &  2 & English \\
   3 & History           &  3 & History           &  6 & History \\
   2 & Philosophy        &    &                   &  2 & Philosophy \\
   7 & Political Science &  6 & Political Science & 13 & Political Science \\
   9 & Sociology         & 29 & Sociology         & 38 & Sociology \\
   9 & Spanish           & 20 & Spanish           & 29 & Spanish \\

\enddata
\end{deluxetable}

\begin{deluxetable}{rlrlrl}
\tablecolumns{6}
\small
\tablewidth{0pt}
\tablecaption{Underachievers}
\tablehead{
\colhead{} &
\colhead{Male} &
\colhead{} &
\colhead{Female} &
\colhead{} &
\colhead{Total} \\
}
\startdata

   5 & Biology           & 4 & Biology           &   9 & Biology \\
   2 & Chemistry         & 1 & Chemistry         &   3 & Chemistry \\
   1 & CIS               &   &                   &   1 & CIS \\
  20 & Economics         & 1 & Economics         &  21 & Economics \\
   4 & English           & 3 & English           &   7 & English \\
   7 & History           & 2 & History           &   9 & History \\
   6 & Mathematics       &   &                   &   6 & Mathematics \\
   1 & Philosophy        & 1 & Philosophy        &   2 & Philosophy \\
   5 & Physics           & 1 & Physics           &   6 & Physics \\
  14 & Political Science & 3 & Political Science &  17 & Political Science \\
   4 & Sociology         & 2 & Sociology         &   6 & Sociology \\

\enddata
\end{deluxetable}

\section{Evidence for Cognitive Thresholds} 

One might reasonably associate mastery of a subject with GPA $>$ 3.5 --
roughly, the minimum threshold to be admitted to graduate school (i.e.,
students who earn equal numbers of A's and B's in their upper division core
courses are borderline for most graduate programs). We find that in Physics
and Mathematics no student with SAT$_{M}$ less than roughly 600 was able to
attain this level of mastery.

Figure 11 shows the upper division in-major GPAs of UO physics and math
graduates from a 5 year period. The math GPAs were computed using a
specific set of rigorous courses taken by graduate school bound majors.
The figure provides at least modest evidence for a minimum cognitive
threshold required for physics and mathematics. That is, the a priori
probability that a student with SAT$_{M}$ $<$ 600 will perform well enough
to be admitted to graduate school in these subjects is extremely low. Note
many poorly performing students switch majors, and hence do not populate
the lower left corner of our graph.

A simple but plausible model for college performance includes (at least)
two factors: 1. ability (measured by SAT) and 2. conscientiousness or
effort (for simplicity, an uncorrelated random variable, probably normally
distributed). In one version of this model the predicted GPA might depend
linearly on the sum of the two factors, each measured in standard deviation (SD) units. (This is
similar to the GPA predictor depicted in Figure 7.) Applying such a model
to our data, we would conclude that even a student in the, e.g., 90th
percentile of work ethic has a low probability of attaining mastery if
their SAT$_{M}$ score is below 600.

To reiterate, SAT$_{M} \approx$ 600 seems to be the lowest score at which
even a very motivated student has a chance for mastery. From the data one
might guess that only for SAT$_{M}$ well above 700 do students have more
than a 50 percent chance of obtaining GPA $>$ 3.5. That is, a student with
average motivation or conscientiousness probably needs SAT$_{M}$ well above
700 to have a high probability of obtaining mastery.

We were unable to find any similar threshold (either in SAT$_{R}$ or
SAT$_{M}$) in other majors, including economics, sociology, history,
philosophy, biology, chemistry, etc.  For example, Figure 12 is the
analogous plot for English (black) and History (blue) majors versus
SAT$_{R}$ (verbal). If a threshold exists it is probably at SAT$_{R}$ of
450 or so.

For the total SAT-taking population, an SAT$_{M}$ score of 600 is about 75th
percentile.  For the overall population, it might be roughly 85th percentile (the result depends on 
assumptions concerning the pool of test takers versus the general population).
SAT$_{M}$ of 750 is roughly 98th percentile for the total SAT-taking
population.

\begin{figure}
\centering
\includegraphics[scale=0.8]{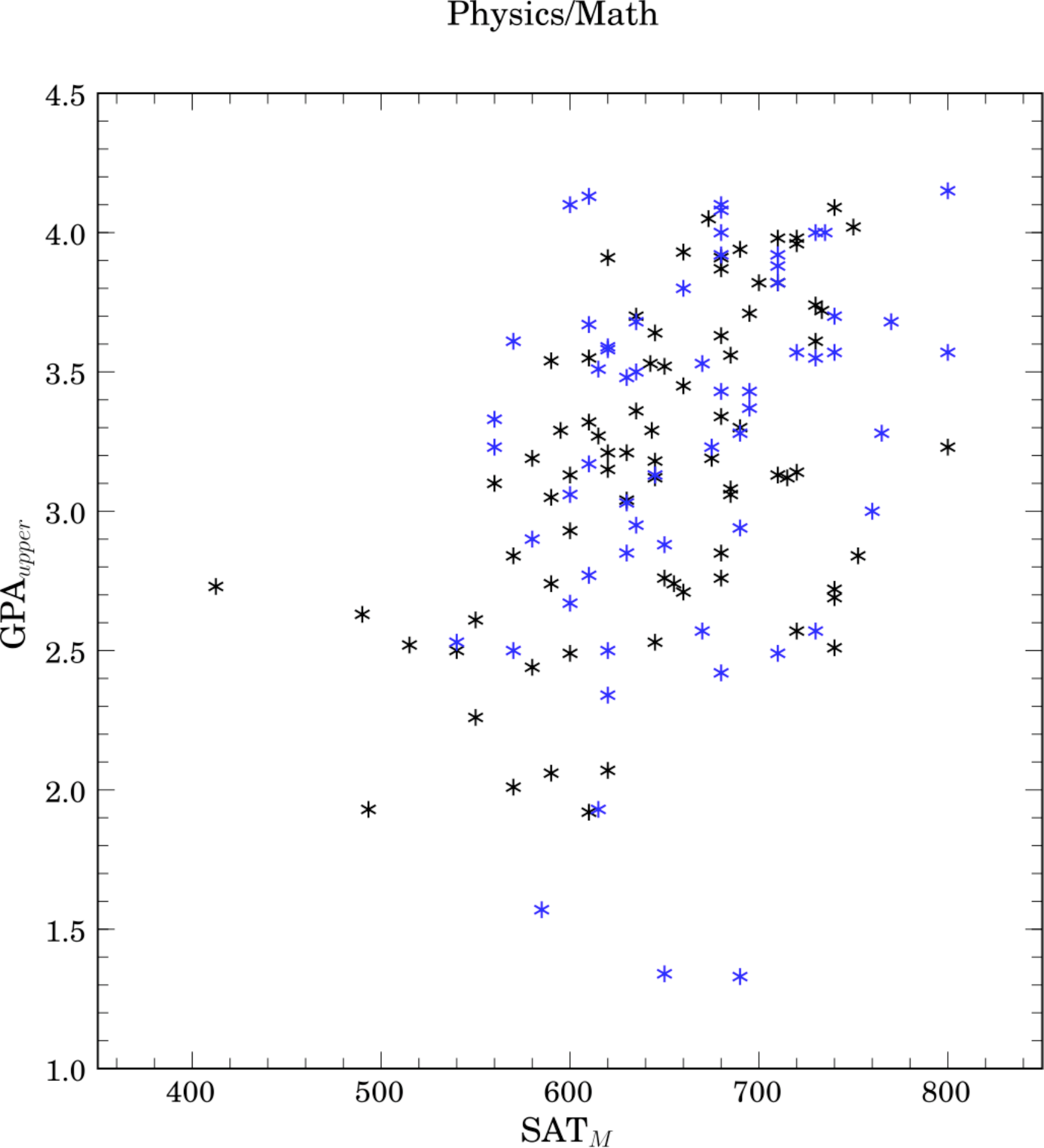}
\caption{Modest evidence for a cognitive threshold is found in the SAT
Math scores of Physics and Mathematics majors.  The black symbols are for
Physics majors, blue for Mathematics.}
\end{figure}

\begin{figure}
\centering
\includegraphics[scale=0.8]{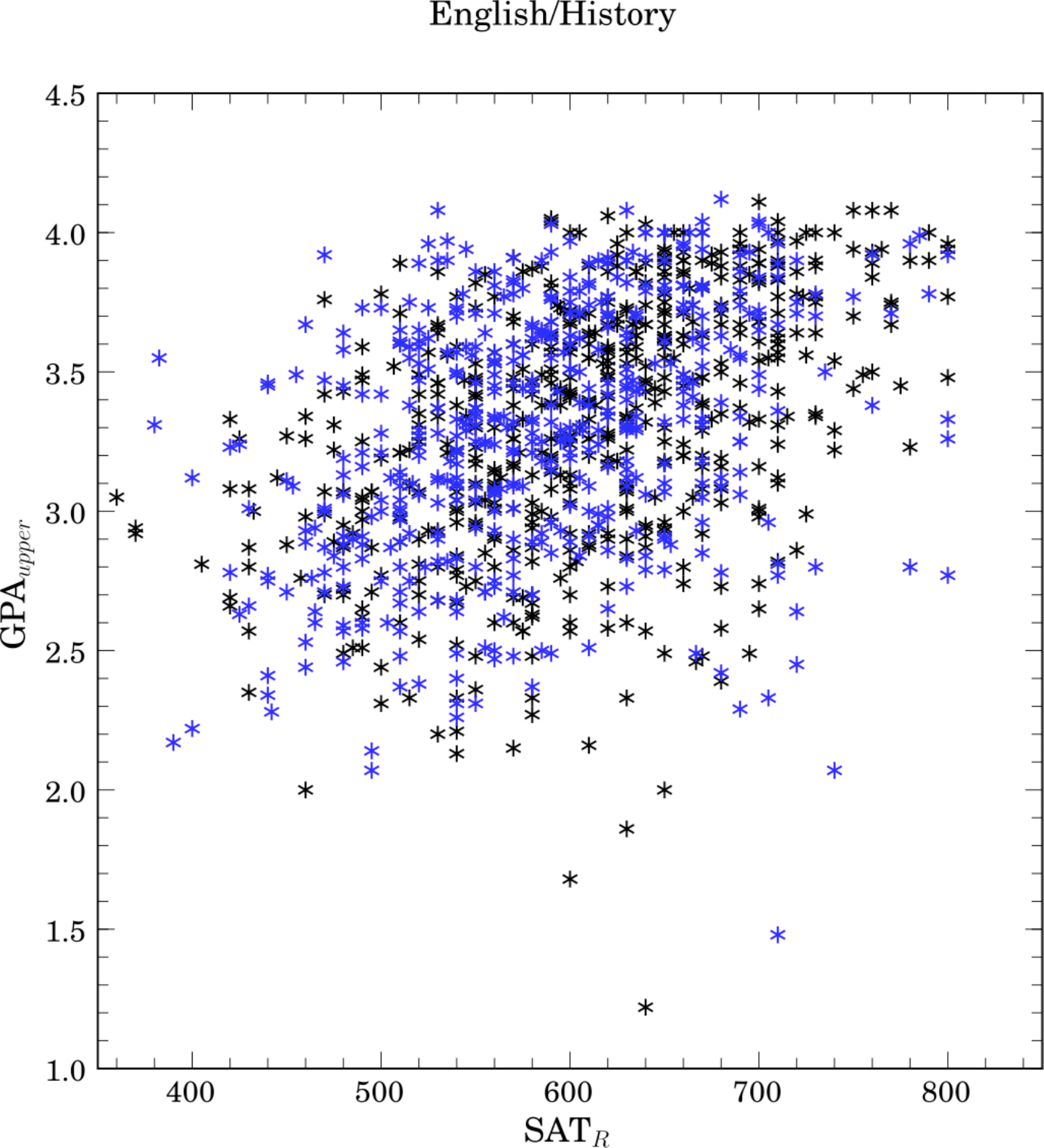}
\caption{In contrast to Figure 11, there is no evidence of a threshold for
English (black) or History (blue) versus SAT$_{R}$.}
\end{figure}

\section{Clark Honors College and the elites} 

The University of Oregon is home to Clark Honors College (CHC), one of the
oldest honors colleges in the country. Students must apply separately to
CHC, and admission is competitive: enrolled students have average SAT of
1340 and unweighted high school GPA of 3.9. In terms of selectivity, the
CHC is roughly comparable to Cornell or UC Berkeley. CHC students must
fulfill additional rigorous course requirements beyond those required by
their major. Their willingness to do so suggests that, in terms of drive
and ambition, they are more similar to students at elite universities than
other UO students with similar SAT and high school GPAs.

The performance of CHC students (red dots) relative to the broader UO
population is shown in Figure 13. The average upper GPA of CHC students is
3.7, versus 3.2 for the entire UO population. This graph plausibly
describes how a group of Berkeley or Cornell students might do at Oregon.
It suggests that UO students with high in-major GPAs have subject mastery
similar to the better students at elite universities. Note that this high
GPA population includes many overachievers who entered the university with
low SATs, low high school GPA or even both. These overachieving students
would have had almost zero probability of admission to an elite university,
an illustration of the imperfections of our system of elite higher
education.

Using our data, we can estimate how a population of elite college students
(e.g., from the Ivy League) might perform at a typical state university.
Students from universities at least as selective as the CHC would be
expected to earn an average upper GPA similar to the 3.7 of CHC students.
This suggests that an average grade of B+ or even A- does not constitute
unreasonable grade inflation at an elite university, if the grade averages
are meant to be commensurate (in the performance or subject mastery they
represent) with those at less selective universities. Similarly, one could
argue that public university graduates with high GPAs (e.g., above 3.7)
likely have levels of subject mastery similar to those of elite graduates.

\begin{figure}
\centering
\includegraphics[scale=0.8]{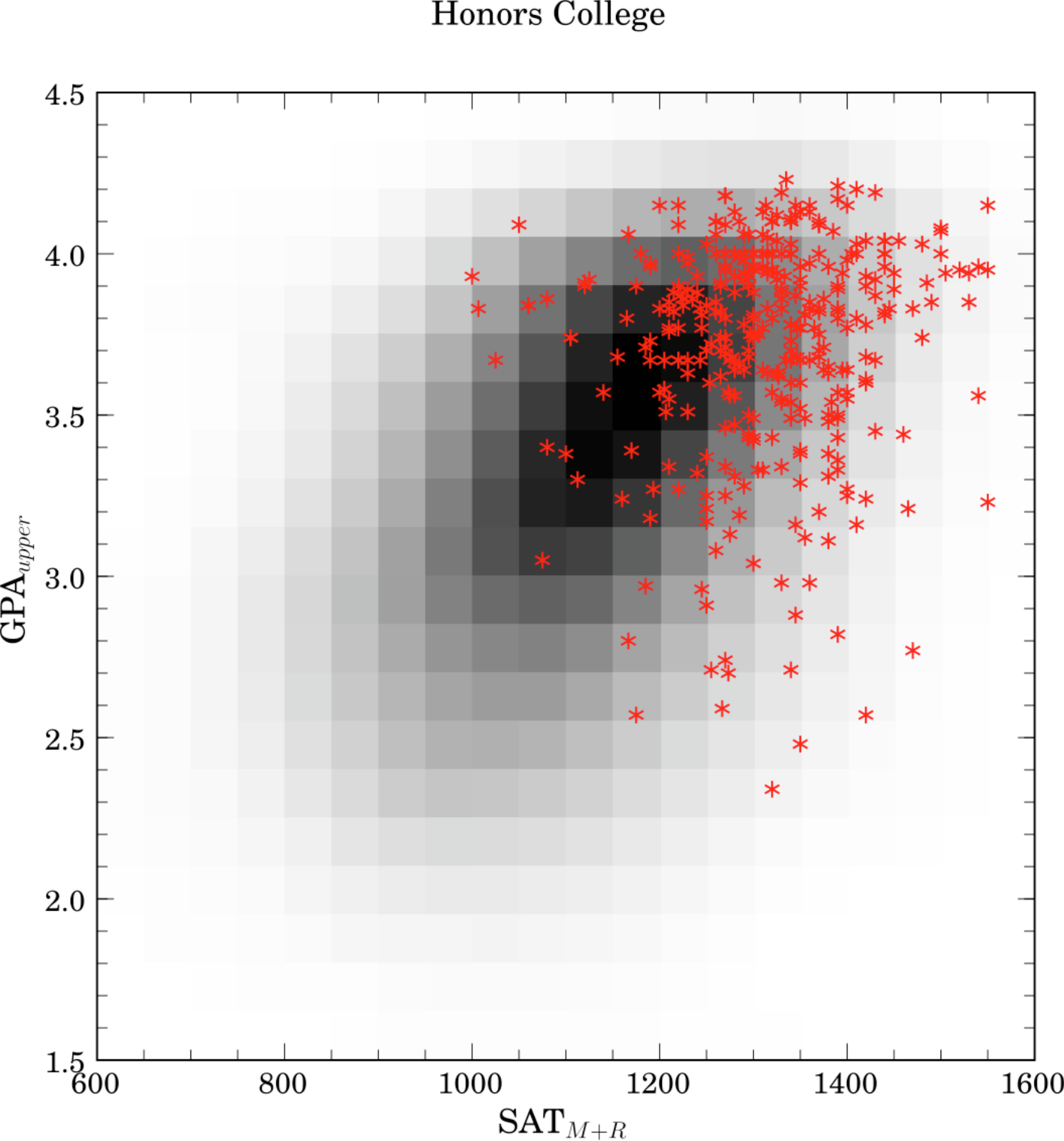}
\caption{A comparison of students from the Clark Honors College with the total student population (restricted to our twelve
CAS majors).}
\end{figure}

\section{Conclusions} 

SAT scores have strong predictive power, if the quantities to be predicted
are carefully defined so as to measure actual academic achievement. The
examples studied in this paper are upper division, in-major GPA and, more
generally, GPA in specific groups of courses, yielding correlations in the
0.35-0.5 range. The often quoted low correlation (roughly 0.3) between SAT
and freshman GPA is subject to self-sorting effects and variation in course
difficulty (students with high academic ability are likely to take more
challenging freshman courses).

The cognitive abilities measured by the SAT seem quite stable: in cases
where GRE scores are available, the correlation between GRE score (measured
4+ years later) and SAT is high (roughly 0.75 for both reading and math).
This is consistent with the results from IQ studies (Jensen 1998).

Below we summarize our most important results.

1. SATs predict upper GPA with correlations in the 0.35 -- 0.50
range.

2. Overachievers exist in most majors, with low SAT scores but
very high GPAs.  These overachievers are disproportionately female.

3. Underachievers exist in all majors, with high SAT scores but
very low GPAs.  These underachievers are disproportionately male.

4. Some majors, like math and physics, may exhibit a cognitive
threshold -- mastery of the material is unlikely below an ability threshold
(as measured by SAT$_{M}$), no matter how hard the student works.

5. Students at public universities, like UO, with high upper GPA
(e.g., 3.7 or greater) likely have subject mastery similar to graduates of
elite universities.  Elite college students who transferred to a state
university would likely average upper division GPAs of 3.7 or greater.

\appendix
\section{APPENDIX}

This study required extensive access to online student records, flexible
network tools and innovative analysis routines.  Almost all the information
used in this project was extracted from student on-line transcripts, but
using automated tools. Without those tools the human labor involved would
have been prohibitive. The data used in our analysis, in particular for
upper division GPA by major, was not available in any SQL database.

Briefly, the data reduction for this study contains three parts.  The first
part is the acquisition of student IDs for the academic terms of interest
in the majors we selected.  This was accomplished using a simple SQL query
to UO's Student Data Warehouse, where we selected all students active in
the Fall terms from 2000 to 2004 for the twelve majors. Selecting by major
lowered the total number of transcripts that had to be queried in the next
stage of the data extraction process.

The second stage was to acquire transcripts, SAT/ACT scores and high school
GPAs from Duckweb servers. Duckweb allows advisors to examine individual
student records using a browser.  In our case, we wanted all the student
information stored on our local machines, so a special set of network tools
was used to extract the relevant pages and download them to a local
machine.  These tools were developed for a NASA data mining project focused
on large online datasets.  Access to modern data repositories is often via
webpages and CGI interfaces.  The tools we used have the ability to behave
like a web browser, but with the advantage of not having to wait for human
input.  They act under automatic control, making decisions based on a set
of user defined conditions.

For Duckweb access, the Python module {\tt urllib} was sufficient to
navigate the series of web pages leading to individual student transcripts
of interest.  Standard CGI security prevents a user from going directly to
a particular webpage (i.e., you cannot navigate directly to the URL
addresses for student records) through the use of cookies.  Fortunately,
{\tt urllib} automatically handles the processing of cookies as long as the
script follows the same path of webpages that an interactive user would.
The primary difference between a script and a human user is, of course, the
speed of interaction with the server.  In order to avoid triggering DOS
(Denial of Service attack) alerts, the scripts were run late at night and
included pause statements in the chain of commands so as not to tax the
servers.  The extraction of 25,000 student records only required an evening
of interaction.  Whereas the scripts could have obtained this information
in only a few tens of minutes, the process was deliberately slowed and took
several hours.

The last stage of our analysis involved parsing the HTML files on each
student into some meaningful dataset (e.g., XML format or raw ASCII).  Once
in a usable format, we continued our analysis using standard tools for
astronomical research (means, correlations, plotting, linear fitting).
Often HTML pages contain embedded information tags which facilitate
converting the pages into XML format.  In Duckweb web data, the HTML tags
are only used for browser formatting.  A simple Python {\tt re} (regular
expression) command strips all the HTML tags from the page.  The remaining
text is simple to process for courses, grades, dates, etc.

All of these steps were accelerated by the Python scripting language.  The
Python language has several advantageous features.  For example, Python's
unique {\tt try/except} feature allows for unknown conditions in the datasets to
be handled without crashing the analysis.  Python's modular nature allows
for network, statistical and plotting features to be seamlessly integrated
without producing an unmanageable amount of code.  In this study, none of
the scripts were over 50 lines in length, giving us the ability to consider
a range of research questions without being limited by pre-existing
software tools.

\acknowledgements

\end{document}